\documentclass[12pt,letterpaper]{article}
\usepackage[top=1in,bottom=1in,left=1in,right=1in]{geometry}
\usepackage[utf8]{inputenc}
\usepackage[dvipsnames]{xcolor}
\usepackage{graphicx}
\usepackage{wrapfig}
\usepackage{boxedminipage}

\usepackage[colorlinks = true,linkcolor = purple,urlcolor  = purple,citecolor = purple,anchorcolor = purple]{hyperref}

\newcommand{\checked}[1]{\textcolor{green}{Checked.}}

\newcommand{\hubunits}{\,{\rm km}\,{\rm s}^{-1}\,{\rm Mpc}^{-1}}
\newcommand{\yr}{\,{\rm yr}}
\newcommand{\arcmin}{\,{\rm arcmin}}

\begin{document}

{\raggedright
\Large
Astro2020 Science White Paper \linebreak

{\bf WFIRST: The Essential Cosmology Space Observatory for the Coming Decade}\linebreak

\normalsize

\noindent \textbf{Thematic Areas:} 7. Cosmology and Fundamental Physics \linebreak

\textbf{Principal Author:}

Name: Olivier Dor\'e
\linebreak						
Institution:  Jet Propulsion Laboratory, California Institute of Technology
\linebreak
Email: olivier.p.dore@jpl.nasa.gov
\linebreak
Phone: 626-375-6347
\linebreak

\noindent
\textbf{Co-authors:} 
\linebreak
C.~Hirata (OSU), Y.~Wang (Caltech/IPAC), D.~Weinberg (OSU), T.~Eifler (U. Arizona), R.~J.~Foley (UCSC), C.~He~Heinrich (JPL/Caltech), E.~Krause (U. Arizona), S.~Perlmutter (UCB/LBNL), A.~Pisani (Princeton), D.~Scolnic (Duke), D.~N.~Spergel (Princeton/CCA), N.~Suntzeff (Texas A\&M), G.~Aldering (LBNL), C.~Baltay (Yale), P.~Capak (Caltech/IPAC), A.~Choi (OSU), S.~Deustua (STScI), C.~Dvorkin (Harvard), S.~M.~Fall (STScI), X.~Fang (U. Arizona), A.~Fruchter (STScI), L.~Galbany (U. Pittsburgh), S.~Ho (CCA), R.~Hounsell (Penn), A. Izard (JPL/Caltech, UC Riverside), B.~Jain (Penn), A. M. Koekemoer (STScI), J.~Kruk (GSFC), A.~Leauthaud (UCSC), S.~Malhotra (GSFC), R.~Mandelbaum (CMU), E.~Massara (CCA), D.~Masters (JPL/Caltech), H.~Miyatake (Nagoya/Kavli IPMU/JPL), A.~Plazas (Princeton), J.~Rhoads (GSFC), J.~Rhodes (JPL/Caltech), B.~Rose (STScI), D.~Rubin (STScI), M.~Sako (Penn), L.~Samushia (KSU), M.~Shirasaki (NAOJ), M.~Simet (UC Riverside, JPL/Caltech), M.~Takada (Kavli IPMU), M.~A.~Troxel (Duke), H.~Wu (OSU), N.~Yoshida (U. Tokyo/Kavli IPMU), Z.~Zhai (Caltech/IPAC)  
}

\bigskip

\noindent
{\bf Abstract:}\\
Two decades after its discovery, cosmic acceleration remains the most profound mystery in cosmology and arguably in all of physics.  Either the Universe is dominated by a form of dark energy (DE) with exotic physical properties not predicted by standard model physics, or General Relativity (GR) is not an adequate description of gravity over cosmic distances.  WFIRST emerged as a top priority of Astro2010 in part because of its ability to address the mystery of cosmic acceleration through both high precision measurements of the cosmic expansion history and the growth of cosmic structures with multiple and redundant probes.  \emph{We illustrate in this white paper how mission design changes since Astro2010 have made WFIRST an even more powerful dark energy facility \cite{[4]}
and have improved the ability of WFIRST to respond to changes in the experimental landscape. WFIRST is the space-based probe of DE the community needs in the mid-2020s.}


\pagenumbering{gobble}
\newpage
\pagenumbering{arabic}

WFIRST, in part because of its unique ability to tackle the mystery of DE with multiple robust probes, became a top priority for Astro2010. The scientific landscape has greatly evolved since Astro2010, and so have the observatory capabilities. Because of its unprecedented efficiency and flexibility as a DE observatory, and because of the exquisite control of systematic effects it enables -- unmatched in any planned mission, WFIRST will address in a definite manner the driving DE question(s) of the next decade. 

\vspace{-0.5cm}
\section{The Dark Energy Landscape in the Mid-2020s}
\vspace{-0.25cm}
\paragraph{A decade rich in observational progress.}Measurements of cosmic expansion and growth of structure have sharpened considerably since Astro2010, thanks to more extensive surveys and improved methodologies \cite{[1]}. Independent measurements from the cosmic microwave background (CMB) have also become much more stringent. The expansion history measurements using Type Ia supernovae (SN) and baryon acoustic oscillations (BAO) has now been measured past $z\sim 1$, and for z$<$1, we have measurements better that one percent.  Measurements of dark matter clustering are now approaching 2-3\% precision, thanks mainly to large weak lensing (WL) experiments such as the Kilo-Degree Survey (KiDS), the Dark Energy Survey (DES), and the Subaru Hyper-Suprime Camera (HSC).
\vspace{-0.5cm}
\paragraph{Tensions in contemporary cosmology.}Most measurements of expansion history are in good experimental agreement with the predictions of a CMB-normalized $\Lambda$CDM cosmological model.  However, most (though not all) WL-based measurements imply an amplitude of low redshift matter clustering smaller than $\Lambda$CDM predictions; the discrepancy with any individual measurement is typically only 1--2$\sigma$, but the collective tension is significant {\em if} one takes reported error bars at face value and treat them as independent \cite{[10]}.  
Growth rate measurements from galaxy redshift-space distortions (RSD) are currently less precise, with 5--10\% errors, so they do not yet help in assessing this tension.  Another significant tension in contemporary cosmology is between direct distance-ladder measurements of the Hubble constant and $\Lambda$CDM predictions constrained by the CMB and BAO, e.g., $H_0 = 73.24 \pm 1.74\hubunits$ \cite{[2]} compared to $67.66 \pm 0.42\hubunits$ \cite{[3]}.  The matter clustering tension could be a signature of modified gravity that alters the rate of low-$z$ structure growth, while the $H_0$ discrepancy could be explained by early DE or additional relativistic species that change the pre-recombination expansion rate and thereby rescale the cosmic ruler imprinted by BAO \cite{[7]}.  Alternatively, they could result from yet undiscovered astrophysical or observational systematic errors. Finally, expansion rate measurement through BAO in the Lyman-$\alpha$ forest at $z\sim$ 2.4, the only high redshift BAO measurement so far besides the CMB, is about 2.4$\sigma$ lower than the expected value given the CMB \cite{[8]}. This unresolved situation provides key lessons for the next decade: \emph{(1) we will need not just larger sample sizes, but also much better control and understanding of systematic errors; (2) we will need multiple and flexible DE probes and surveys to enable cross-validation and reach robust conclusions.} 
\vspace{-0.5cm}
\paragraph{Anticipating the coming decade.} By the time WFIRST launches in the mid-2020s, we will have the final results from DES, KiDS, and HSC, most of the BAO and RSD results from the Dark Energy Spectroscopic Instrument (DESI), and early data from the Large Synoptic Survey Telescope (LSST) and the Euclid mission.  Anticipating these results, we can imagine three somewhat different scenarios. (1) If the central values of cosmological measurements remain as they are and uncertainties shrink, then the $\Lambda$CDM model will be ruled out, and by 2025 we will be deep into trying to understand the alternatives, e.g., by mapping out the redshift history of deviations in the expansion rate or the rich phenomenology of modified gravity models, in which deviations from GR-based clustering can be scale, environment, and tracer dependent. The combination of DE probes and redshift range enabled by WFIRST in combination with these surveys will be invaluable. (2) Alternatively, the current tensions may have dissipated and been replaced by new ones at the $\sim 3\sigma$ level of the much reduced statistical errors. In this case, the crucial task for LSST, Euclid, and WFIRST will be to verify these tensions or show that they arise from statistical flukes or underestimated systematics. (3) Finally, results in the mid-2020s could all be consistent with $\Lambda$CDM, and the collective power of LSST, Euclid, and WFIRST will be needed to reach another factor of ten into cosmological parameter space. \emph{These scenarios highlight that enabling multiple and flexible DE probes should be an essential requirement for the WFIRST DE program. We now discuss how the current design achieves this goal.}

\vspace{-0.5cm}
\section{WFIRST: An Agile and Efficient DE Observatory}

The observational situation summarized above, and analogous challenges in the search for primordial gravitational waves, motivate WFIRST DE capabilities design philosophy. \emph{WFIRST DE capabilities are designed (1) to allow multiple DE probes to increase the interpretation robustness, (2) to maximize systematics error control and not only statistical power, (3) to enable different surveys to best address the relevant questions in the next decade.}  If we make a statistically significant detection of deviations from GR or complex dynamics of DE, we want to be able to convince ourselves and the community that this is compelling evidence of new physics, not an artifact of instrumental or astrophysical systematics.

The current design of WFIRST is summarized in \cite{[4]}, which emphasizes ability to address a wide range of scientific questions. WFIRST is a 2.4-m telescope equipped with a coronograph and a wide field instrument (WFI) comprised of 18 4k$\times$4k near-IR detectors (H4RGs) with $0.11^{\prime\prime}$ pixels and a $0.282\deg^2$ field of view.  The filter wheel contains seven filters spanning 0.48-2.0\ $\mu$m, a spectral resolution $R = 435-865$ grism for the slitless galaxy redshift survey, and an $R = 70-140$ prism for slitless spectroscopy of supernovae and their host galaxies. The primary mission duration is 5 years, with a goal of 10 years.
\vspace{-0.5cm}
\paragraph{Unprecedented systematic error control for WL, SN, BAO, and RSD.} WFIRST has the advantages of high angular resolution, high near-IR sensitivity, and high photometric stability, all uniquely enabled by a space-based observatory. Relative to the H2RG detectors on JWST and Euclid, the H4RGs have much improved persistence performance and will use readout electronics with significantly better noise properties. The optical chain is designed for nanometer level stability over the course of an exposure, and the internal calibration system is designed to provide multiple cross-checks of the non-linearity via in orbit calibration. Unlike Euclid, WFIRST has a four band strategy for the WL survey, i.e., galaxy shapes will be measured with high signal to noise in four bands, and allow internal checks from cross-correlations between bands, in addition to richer information on the source and lens galaxies. The galaxy redshift survey is smaller in area than that of Euclid but substantially deeper, with multiple dither and roll angles to maximize uniformity.  Control of astrophysical systematics comes partly from cross-checks among three different approaches to distance and structure growth measurements.  There are also cross-checks internal to each method, e.g., with different classes of SNe or host galaxies in the SN survey, with the complementary cosmological sensitivity of cosmic shear, galaxy-galaxy lensing, and cluster lensing in the WL survey, and with higher order clustering statistics and subdivisions of the galaxy population enabled by the high density of the redshift survey.  Finally, another critical aspect of systematics control is that the reduced and calibrated WFIRST data will be made public immediately, and the refined data sets used for cosmological analyses will be made public as the analyses themselves are completed.  Thus, every cosmological measurement can be scrutinized and repeated by multiple independent teams with distinct approaches. 
\vspace{-0.5cm}
\paragraph{An efficient and flexible cosmological observatory.} WFIRST is a highly efficient facility for cosmological surveys.  For example, in an H-band only survey, WFIRST could image sky to 26.9 mag AB  at a rate of $\sim 8,000 \deg^2/\yr$, well matched to LSST and $\sim 3$ mag deeper than Euclid H-band.  WFIRST is thus well positioned to respond to changes in the experimental landscape and to the performance and results of LSST and Euclid, by adjusting the survey strategy parameters (depth, area, and number of bands) and priorities among the DE surveys \cite{[15]}.  For example, if other observations highlight expansion history at $z=1-2$ as a critical area, then WFIRST could prioritize high-redshift SNe and the galaxy BAO survey.  Conversely, if observations provide tantalizing evidence for modified gravity, then WFIRST could prioritize deep WL observations or an H-band survey of the LSST footprint that would yield optimal shape measurements from the combined data set. 

\begin{figure}[t]
\centering
\includegraphics[width=0.99\textwidth]{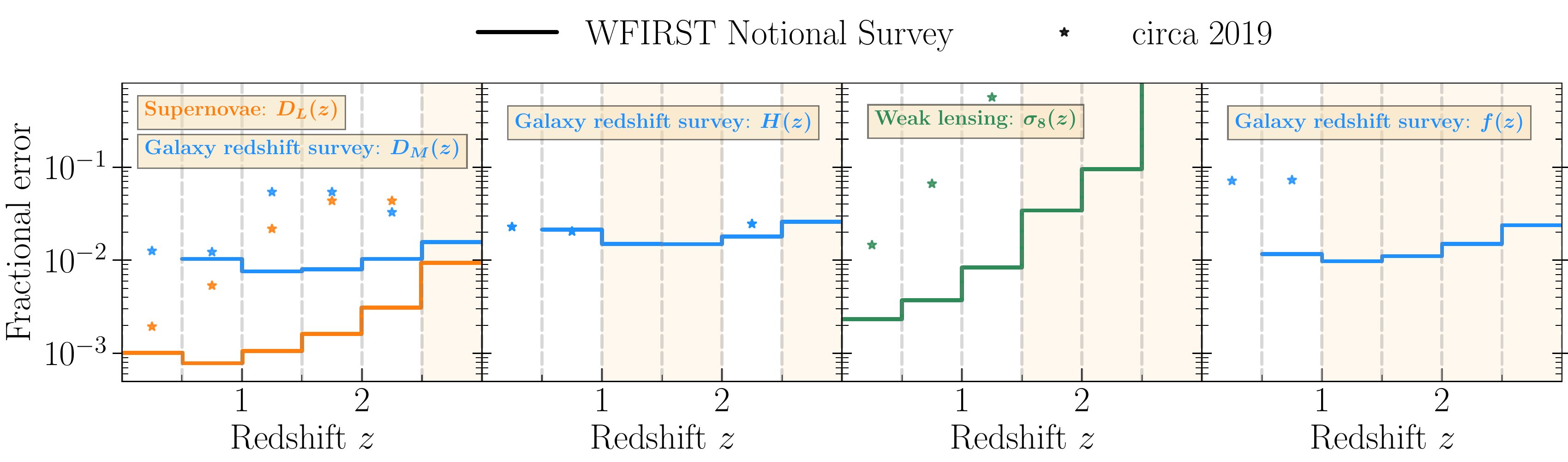}
\vspace{-0.45cm}
\caption{\small{WFIRST notional survey performance for multiple DE probes (solid lines) compared to current measurements (stars). WFIRST will measure the expansion history at sub-percent level over a broad range of redshifts using its SN survey (1st panel from left), the high-latitude spectroscopic survey (first and second panels), the high-latitude imaging survey (third panel), as well as the growth rate $f= d\ln D/da$ where $D$ is the linear growth rate and $a$ is the expansion rate (fourth panel). WFIRST DE probes will open new redshift windows shown by the shaded areas.}}
\label{fig:DE_panel}
\vspace{-0.5cm}
\end{figure}

\vspace{-0.5cm}
\section{WFIRST Notional Cosmological Surveys}

To define the science requirements, the WFIRST formulation teams have designed notional DE surveys.  These would be preceded by pilot surveys early in the mission, to test the performance of WFIRST against pre-flight expectations and provide revolutionary data sets for immediate community use. In our notional surveys the total time for the SN program is 0.5 yrs (spread over the middle 2 yrs), and the full time for the high-latitude survey (HLS) imaging and spectroscopic surveys is 1.05 yrs and 0.59 yrs, respectively (spread over the five year prime mission).
\emph{The actual allocation of observing time will be decided close to launch and revisited during the mission through a peer-reviewed process, and the allotment to DE programs may well differ from 2.14 years. The time spent on cosmology surveys in the prime+extended mission will depend on developments in the field, and the value of these surveys for other areas of astrophysics.}
\vspace{-0.5cm}
\paragraph{WFIRST HLS.} The main trades to consider are area and depth, the number of bands, and the balance between imaging and spectroscopy. The notional HLS covers $2,000~\deg^2$ interleaving imaging and grism spectroscopy over the same footprint.  The observing and dither strategy yields 5--9 exposures of each point in the footprint, with 140 s exposures in each filter and 297 s in the grism.  The expected imaging depth (5$\sigma$ point source AB) is 26.9, 26.95, 26.9, and 26.25 in Y, J, H, and F184 respectively, well matched to the depth of LSST in optical bands given typical galaxy colors. The effective WL source density in the co-added imaging is $50 \arcmin^{-2}$ \cite{[6]}, for a total of $3.6\times 10^8$ shape measurements, corresponding numbers for H-band alone are $35 \arcmin^{-2}$ and $2.5\times 10^8$.   The expected spectroscopic depth (5$\sigma$ point source emission line sensitivity) is $7\times 10^{-17}\,{\rm erg}\,{\rm s}^{-1}\,{\rm cm}^{-2}$ at the center of the $1.00-1.93$ $\mu$m bandpass.   The expected yield and comoving space density of the galaxy redshift survey are more uncertain because of remaining uncertainties in the galaxy emission-line luminosity function.  Our forecasts suggest $n=1.3\times 10^{-3} h^3{\,\rm Mpc}^{-3}$ at $z=1.5$, with $nP_{0.2} = 3.3$ in the sampling figure-of-merit often quoted for BAO analysis, and a total of 14 million H$\alpha$ redshifts ($z=0.52-1.94$) and 3.8 million [OIII] redshifts ($z=1.00-2.85$).  Relative to the Euclid redshift survey, the WFIRST redshift survey will be smaller in area but higher in density.  Our forecasts suggest that these differences approximately cancel for BAO analyses at $z\sim 1.7$.  The high density of the WFIRST sample offers unique opportunities for studying non-linear galaxy clustering at $z=1-2$, with clustering measures like the bispectrum and the abundance and profiles of cosmic voids \cite{[11],[13]}.

\vspace{-0.5cm}
\paragraph{WFIRST SN survey.} Many trades need to be considered, and optimization remains an area of active investigation.  The number of tiers, number and exact filters for each tier, and cadence can be traded with area and depth. The notional SN survey has two reference points: one that exclusively uses the imaging capabilities of the WFI \cite{[5]} and one that uses the prism for half of the SN time.  Both use six months of integrated time spread over two years, with three tiers of depth and area. 
The half-prism-half-imaging survey observes 4.76/1.12/0.84~$\deg^2$ each five days in observer frame, with prism exposure times of 600/3,600/9,000~s respectively, for a total of 15 hr, and the four-filter imaging follows the same cadence, with the same 15 hr total exposure time per visit. This survey would observe $\sim$3,400 SNe~Ia, 1,200--1,600 of which would be spectroscopically sub-typed, and of suitable S/N for spectroscopic population and evolution studies.  Altogether, the half-prism-half-imaging survey should yield more than 100 SN per 0.1 redshift bin with a shallow S/N = 15 spectroscopic time series (suitable for redshift determination) up to a redshift of 1.8 and a deeper ${\rm S/N} = 35$ (suitable for sub-typing and evolution studies) up to a redshift of 1.2. The all-imaging survey (\emph{Imaging:Allz} in \cite{[5]}) consists of three imaging tiers achieving an individual exposure depth (total stacked depth) of 22.3, 24.5, and 26.1~mag (25.0, 27.2, and 28.8~mag) in $RZY\!J$, $RZY\!J$, and $Y\!J\!H\!F$ covering 49, 20, and 9~deg$^{2}$ for each tier, respectively, with a cadence of 5 days.  This survey would measure useful distances to $\sim$11,000 SNe~Ia to $z = 3$. 
The balance between WFIRST SN imaging and spectroscopy will depend on the level of systematics control required in each redshift range, and the ability of the prism, lightcurves, and external sources to provide it.

\vspace{-0.5cm}
\section{Cosmological impact of the WFIRST DE program}

\begin{wrapfigure}{r}{0.45\textwidth}
  \vspace{-2.cm}
  \begin{center}
    \includegraphics[width=0.45\textwidth]{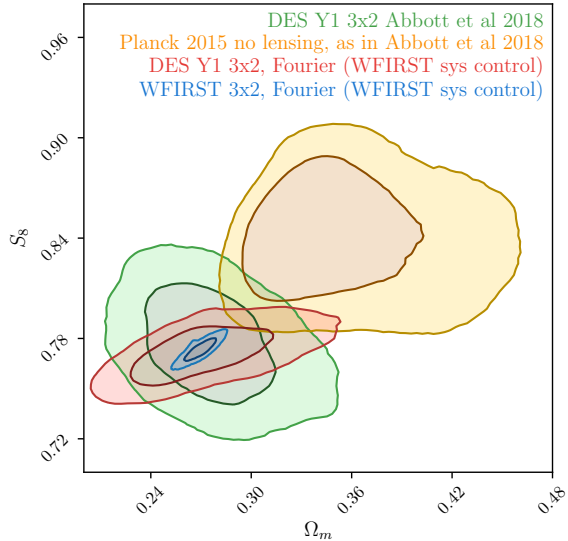}
        \vspace{-2.0cm}
    \caption{\small{Tension between the Planck (yellow contours, 68 and 95\% confidence level) and DES (green) measurements in the $S_8=\sigma_8(\Omega_m)^{0.5}$-$\Omega_m$ plane. Red contours represent constraints from a DES Y1 like experiment if systematic effects were mitigated at the level expected from WFIRST. Blue contours correspond to WFIRST notional survey forecast (imaging only, no BAO, no SN), including the expected level of systematic effects \cite{[6]}.}}
    \label{fig:WF_DES}
    \end{center}
    \vspace{-0.75cm}
\end{wrapfigure}
\vspace{-0.25cm}
\paragraph{Robustness and discovery space.} To illustrate the power and complementarity of the multiple WFIRST DE probes, we illustrate in Fig.~\ref{fig:DE_panel} how WFIRST notional surveys will achieve multiple (sub)percent accuracy measurements of the expansion history of the Universe as a function of redshift, as well as on the structure growth rate. This combination of probes will allow multiple cross-checks and enable a robust interpretation. 

Because WL depends on both the cosmic expansion history and the growth of matter clustering, Fig.~\ref{fig:DE_panel} does not fully capture its impact. Fig.~\ref{fig:WF_DES} illustrates the $S_8$ tension mentioned above that shows tantalizing evidence for departure from $\Lambda$CDM. The blue countour shows how WFIRST imaging data alone could illuminate it. This figure also illustrates the importance of controlling systematic effects in WL surveys. Both figures highlight that the potential discovery space for WFIRST is enormous --- there is lots of room for deviations from GR+$\Lambda$CDM that would be consistent with current data and robustly detected by WFIRST.

\vspace{-0.5cm}
\paragraph{Synergies with Other Surveys.} 
As highlighted by \cite{[14],[15]}, the cosmological science enabled by the combination of WFIRST, LSST, and Euclid goes well beyond what any of these projects can achieve alone, especially when
it comes to controlling systematic uncertainties and model 
degeneracies. Combining WFIRST and LSST in particular will be critical for reliable photometric redshifts (object by object or in cross-correlation), source deblending, galaxy shape measurement validation, and modeling at small physical scales.


The WFIRST [O{\,\sc iii}] survey provides a probe of the high-redshift Universe that is independent of and complementary to that of the DESI Lyman-$\alpha$ forest survey, including growth of structure constraints from galaxy RSD. There is a natural synergy between the (Southern) WFIRST SN survey and the LSST deep-drilling fields: they should overlap and have similar depth observations at the same epochs during the two-year WFIRST SN survey \cite{[12]}.  This strategy will increase the wavelength range of the SN observations to cover 0.3--2.0~$\mu$m, which extends the redshift range for which SNe have similar rest-frame observations.  Such a combined survey will provide additional systematic cross-checks ranging from discovery to photometry (the bluer WFIRST filters overlap with the redder LSST filters) to SN systematic uncertainties related to dust to spectro-photometric control of SN population-drift systematics.

\newpage

\section*{Acknowledgements}
Part of this research was carried out at the Jet Propulsion Laboratory, California Institute of Technology, under a contract with the National Aeronautics and Space Administration. \copyright 2019. All rights reserved.







\end{document}